\documentstyle[prb,aps,floats,psfig]{revtex}

\begin{document}
\twocolumn[
\hsize\textwidth\columnwidth\hsize\csname@twocolumnfalse\endcsname

\title{Theory of the $c$-Axis Penetration Depth in the Cuprates}
\author{R. J. Radtke and V. N. Kostur\cite{na}}
\address{Center for Superconductivity Research, Department of Physics,\\
University of Maryland, College Park, Maryland, 20742-4111}
\author{K. Levin}
\address{Department of Physics and The James Franck Institute,\\
The University of Chicago, Chicago, Illinois  60637}
\maketitle

\begin{abstract}
Recent measurements of the London penetration
depth tensor in the cuprates find a weak
temperature dependence along the $c$-direction which
is seemingly inconsistent with evidence for $d$-wave pairing
deduced from in-plane measurements.
We demonstrate in this paper that these disparate results are
not in contradiction, but can be explained within a theory based
on incoherent quasiparticle hopping between the CuO$_2$ layers.
By relating the calculated temperature dependence of the
penetration depth $\lambda_c(T)$ to the $c$-axis resistivity, we
show how the measured ratio $\lambda_c^2 (0) / \lambda_c^2 (T)$
can provide insight into the behavior of $c$-axis transport below
$T_c$ and the related issue of ``confinement.''
\\
\end{abstract}

]

Measurements of the temperature dependence of the in-plane
penetration depth in $\rm YBa_2Cu_3O_{7-\delta}$ (YBCO) have
been interpreted as providing strong support for a $d$-wave
order parameter.\cite{Hardy}
With the availability of high-quality single crystals, these
measurements are now being extended to the $c$-axis direction in
YBCO\cite{Bonn,Anlage} as well as in
$\rm Bi_2Sr_2CaCu_2O_8$ (BSCCO)\cite{KM} and
$\rm La_{1-x}Sr_xCuO_4$ (LSCO).\cite{Shibauchi}
There appears to be one consistent feature of these different
experiments:  the (low) temperature dependence of the $c$-axis
penetration depth $\lambda_c (T)$ is much weaker than that
observed in the $ab$ plane.
In addition, the penetration depth ratio
$\lambda_c^2 (0) / \lambda_c^2 (T)$ in YBCO is linear in $T$ at
low $T$ with a small slope that decreases with decreasing oxygen
content.\cite{Bonn}

The present paper addresses these penetration depth data
in conjunction with the $c$-axis resistivity $\rho_c (T)$
for a variety of different cuprates.
We argue that the temperature dependence of $\rho_c (T)$
essentially determines that of $\lambda_c (T)$; in particular,
$\lambda_c (T)$ is expected to have a weak temperature dependence
at low T whenever $\rho_c (T)$ exhibits a strong semiconducting
behavior.
In this way, we can reconcile the $ab$-plane data supporting
$d$-wave pairing with the observed $c$-axis behavior.
In addition, by presenting typical results for $\lambda_c$
with the corresponding $\rho_c$, we illustrate the
generic features of these quantities without relying on
specific parameterizations of (or fits to) existing data.
Through the study of the $c$-axis coupling, we touch on the
issues which may lie at the heart of the nature of the normal
state\cite{AZ} and also potentially the mechanism
of high-temperature superconductivity.\cite{AC}
In particular, our results imply that a Fermi-liquid-based
description of the electronic states in the CuO$_2$ planes
appears as consistent with the data as theories based on
the idea of ``confinement,'' even though the underlying
assumptions are considerably different.

To understand the behavior of $\lambda_c (T)$ and the nature
of $c$-axis coupling below $T_c$, we utilize an incoherent
hopping model for the $c$-axis coupling.\cite{Rojo,RLL,RL,RK}
Above the critical temperature $T_c$, this model views each CuO$_2$
layer as a two-dimensional Fermi liquid which is weakly coupled
to its nearest neighbors.
This model yields a semiconducting $T$ dependence of
$\rho_c (T)$ ($d\rho_c (T) / dT < 0$) while maintaining a finite
residual resistivity $\rho_c (0)$.
The semiconducting behavior is also associated
with the absence of a Drude peak
in the $c$-axis optical conductivity.\cite{Rojo}
Thus, while the $c$-axis transport properties suggest
an insulating state at low temperatures, the actual
zero-temperature state is nevertheless metallic.\cite{Rojo}
Extending this model to $\lambda_c$ and comparing
its predictions with the available experimental data provides
an opportunity to learn about the precise low-$T$ behavior of the
$c$-axis resistivity as well as the degree to which
the layers communicate in the superconducting phase.

We begin by writing the Hamiltonian for the electronic system
as the sum of a Hamiltonian for the individual CuO$_2$ layers,
$\sum_m H_m$, and an interlayer coupling term $H_{\bot}$:
$H_{\rm el} = \sum_m H_m + H_{\bot}$.
We leave the intra-layer Hamiltonian unspecified except to
demand that each $H_m$ yield a two-dimensional Fermi liquid
which becomes a either an $s$- or a $d$-wave superconductor
below a critical temperature $T_c$.
$H_{\bot}$ is taken to be\cite{Rojo,RLL,RL,RK}
\begin{equation}
H_{\bot} = \sum_{im\sigma} t_{im}
  c_{i,m+1,\sigma}^{\dag} c_{im\sigma}^{ } + {\rm h.c.},
\label{eq:Hbot}
\end{equation}
where $t_{im} = t_{\bot} + V_{im} + \sum_{j} g_{i-j,m}\phi_{jm}$
and $c_{im\sigma}$ is the usual quasiparticle annihilation
operator for site $i$ in layer $m$ and spin projection $\sigma$.
Physically, the interlayer coupling arises from quasiparticle hopping
due to wave-function overlap (parameterized by $t_{\bot}$),
impurity scattering (modeled by the random variable $V_{im}$),
and bosonic scattering (represented by the field $\phi_{im}$ which
couples to the electronic quasiparticles with strength $g_{im}$).

In treating $H_{\bot}$, we are guided by two observations:
(1)~the mean free path in the $c$-direction extracted from
normal-state transport measurements is less than the lattice
spacing,\cite{Leggett,Cooper} and
(2)~the $c$-axis properties in the superconducting state
are consistent with a picture where nearest-neighbor CuO$_2$
layers form an SIS tunnel junction.\cite{KM,Cooper,Gray}
These observations suggest that the $c$-axis transport may be
viewed as an incoherent tunneling process.
Several theories of this incoherence
exist,\cite{AC,Rojo,Leggett,Graf}
but for simplicity we adopt the phenomenological model of
Ref.~\onlinecite{Rojo} and simulate the effects of incoherence
by performing our calculations to second order in $H_{\bot}$.
This procedure immediately yields an intrinsic
Josephson effect\cite{RL} and a reasonable magnitude and
$T$ dependence of the $c$-axis
resistivity.\cite{Rojo,RLL,RL,note1}

\begin{figure}[tb]
\centerline{
\psfig{figure=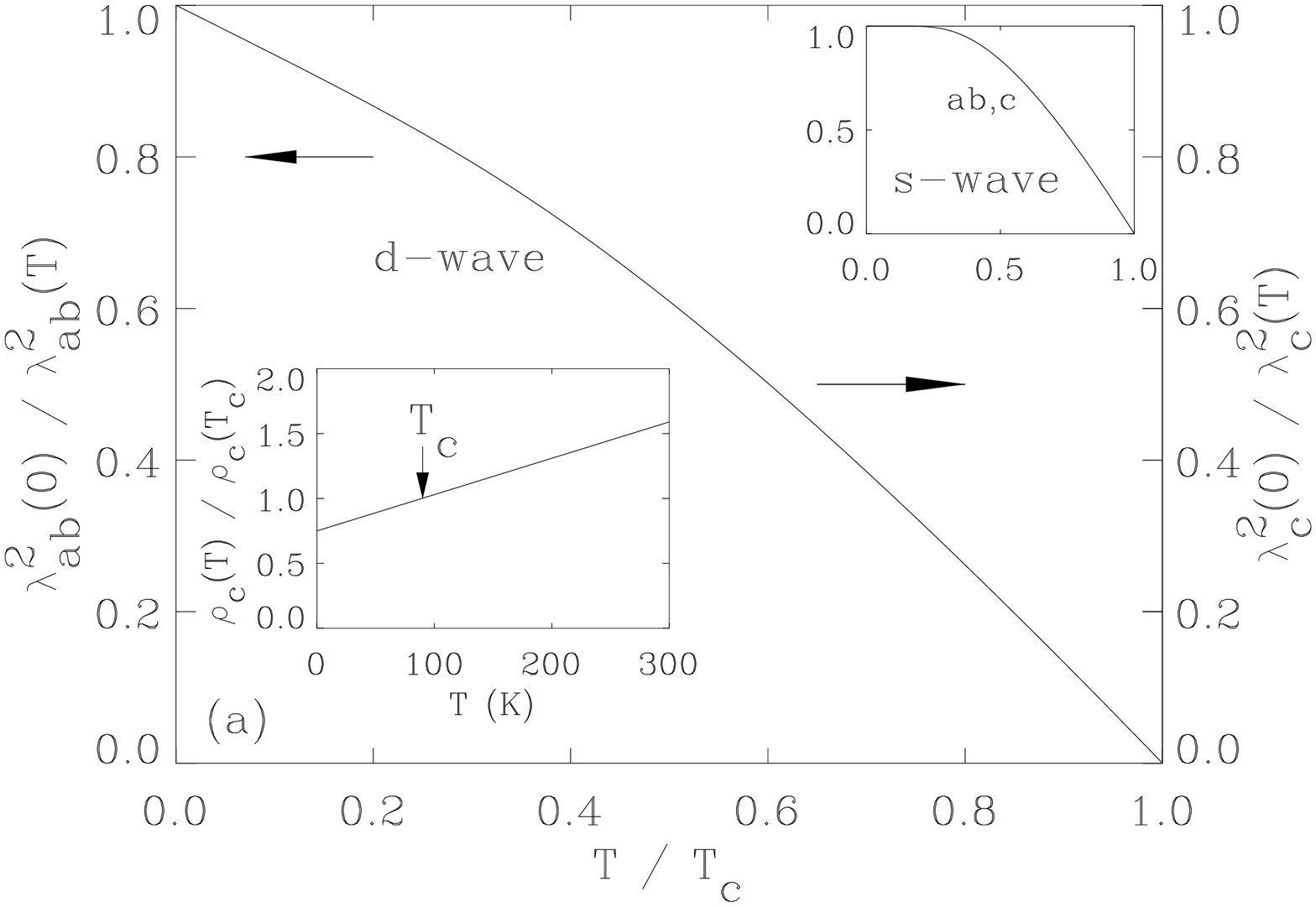,width=0.95\linewidth,angle=90}}
\centerline{
\psfig{figure=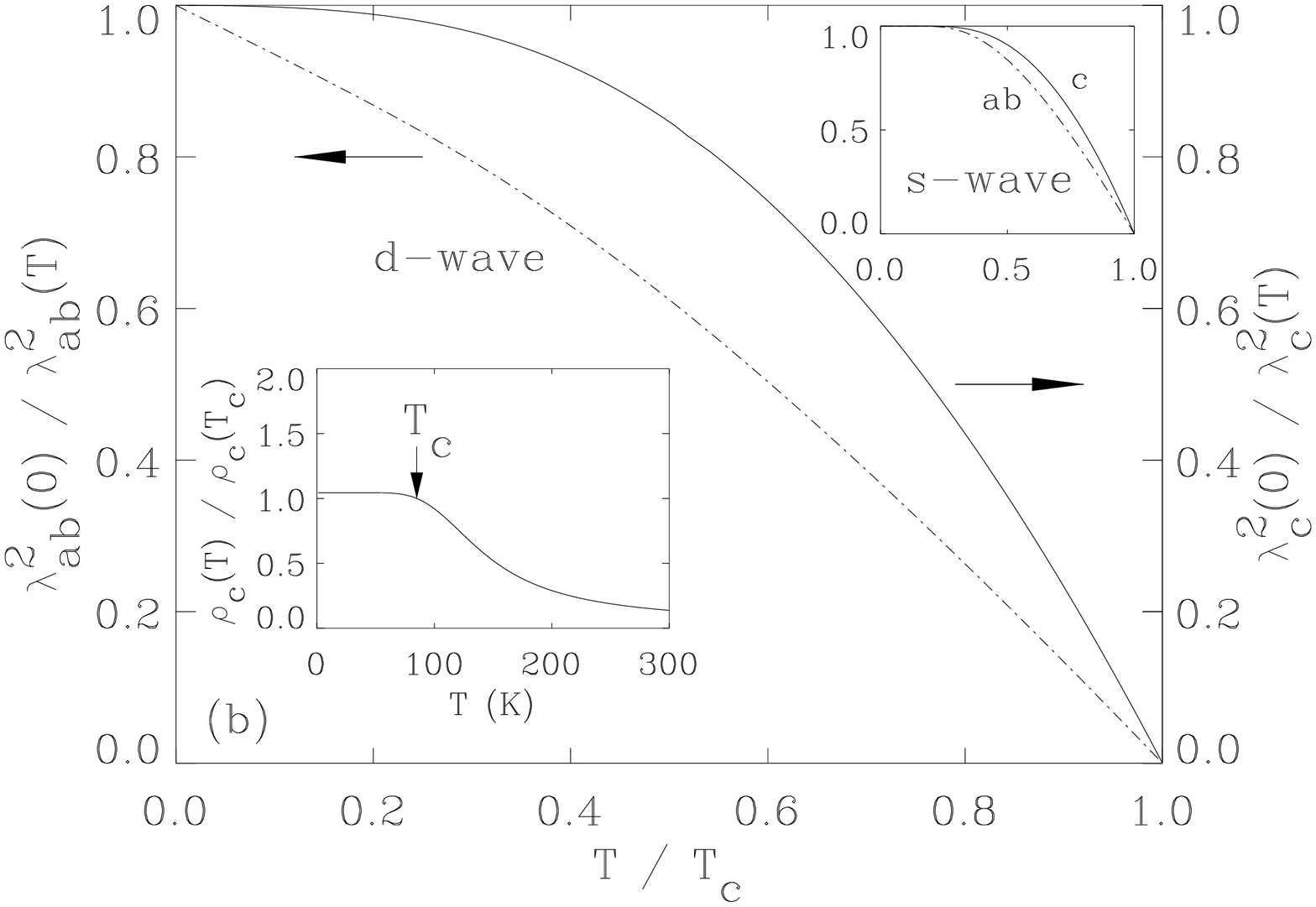,width=0.95\linewidth,angle=90}}
\centerline{
\psfig{figure=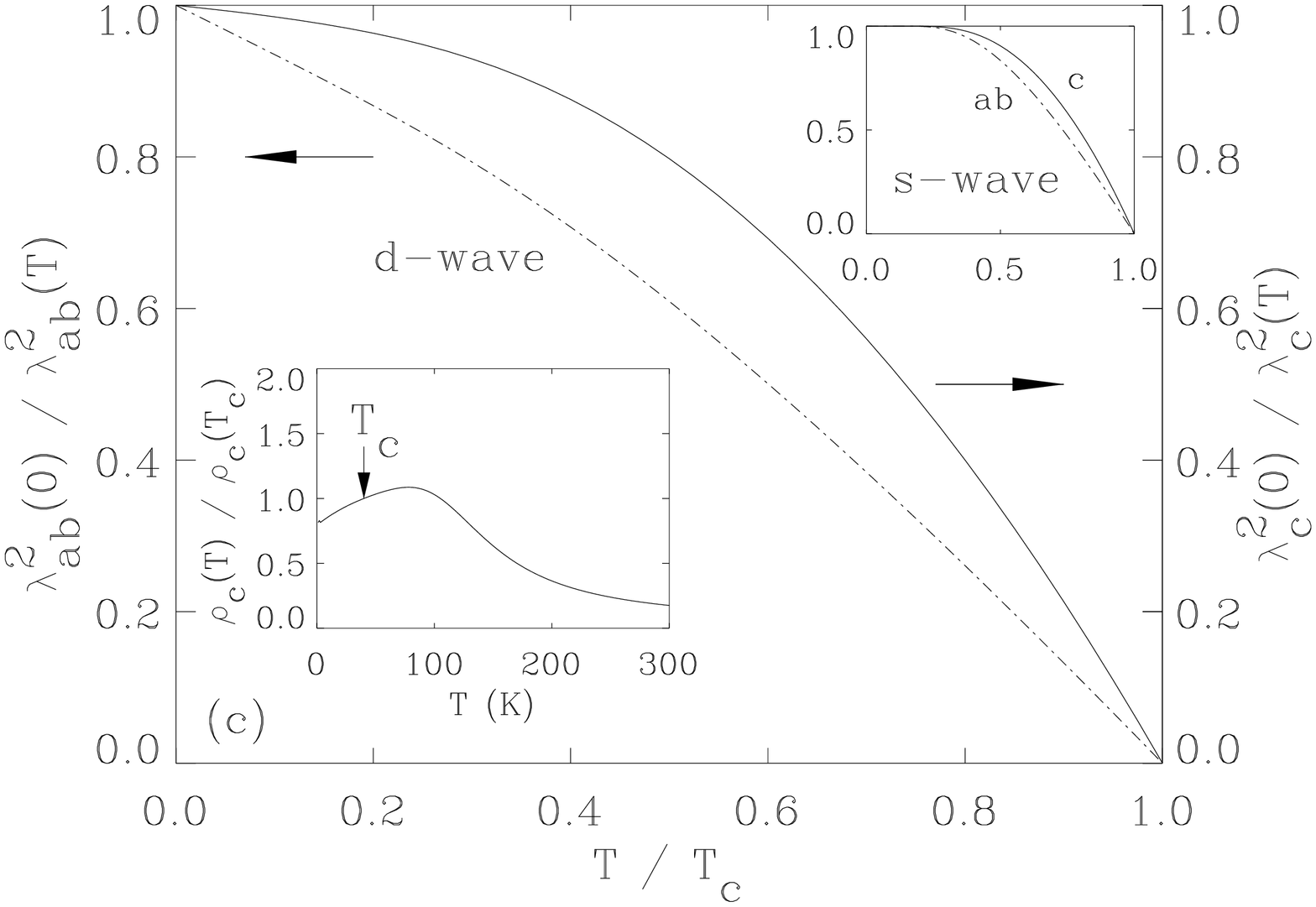,width=0.95\linewidth,angle=90}}
\caption{
Penetration depth ratios $\lambda_{ab}^2 (0) / \lambda_{ab}^2 (T)$
(solid line) and $\lambda_{c}^2 (0) / \lambda_{c}^2 (T)$
(dot-dashed line) as a function of the normalized
temperature $T / T_c$ for both $d$-wave
(main figure) and $s$-wave (upper right inset) pairing
in different limits of the incoherent hopping model
described in the text.
The corresponding $c$-axis resistivity $\rho_c$ is shown
in the lower-left inset normalized to its value at $T_c$
as a function of temperature.
The limits illustrated are
(a)~direct hopping, which may be relevant for
$\rm YBa_2Cu_3O_{6.9}$,
(b)~assisted hopping, which may be the case in
$\rm Bi_2Sr_2CaCu_2O_8$, and
(c)~a combination of these processes, which may obtain
in underdoped compounds like $\rm YBa_2Cu_3O_{6.4}$.
Note in (a) that both the $ab$- and $c$-axis penetration
depth ratios are the same when only direct hopping
is present.}
\label{fig:lambdac}
\end{figure}

In the normal state, we can calculate the $c$-axis dc conductivity
by direct analogy with the problem of tunneling in a normal
metal-insulator-normal metal junction\cite{AZ,RL,LV}
and obtain a $c$-axis conductivity which
is the sum of the conductivities due to each process:
$\sigma_c = \sigma_c^{\rm direct}
  + \sigma_c^{\rm imp} + \sigma_c^{\rm inel}$.\cite{RL}
Generalizing the results of Ref.~\onlinecite{RL} to anisotropic
scattering, we can write these terms as
\begin{equation}
\sigma_c^{\rm direct} = \sigma_0 \, N_0 t_{\bot}^2 \,
  \left( \frac{\tau_{ab}}{\pi \hbar} \right),
\label{eq:o_direct}
\end{equation}
\begin{equation}
\sigma_c^{\rm imp} = \sigma_0 \, N_0^2 \,
  \left< \left< \left| V_{\bf k - k'} \right|^2
  \right>_{\bf k} \right>_{\bf k'},
\label{eq:o_imp}
\end{equation}
and
\begin{equation}
\sigma_c^{\rm inel} = \sigma_0 \, N_0^2
  \left< \left< \left| g_{\bf k - k'} \right|^2
  \right>_{\bf k} \right>_{\bf k'}
  \frac{\hbar \Omega_0 / 2 k_B T}{\sinh^2
  \, (\hbar \Omega_0 / 2 k_B T)}.
\label{eq:o_ba}
\end{equation}
Here, $N_0$ is the density of states per unit cell
per spin at the Fermi surface,
$\hbar / \tau_{ab}$ is the intra-layer scattering rate,
$\sigma_0 = (4 \pi e^2 / \hbar) (d / a^2)$, $d$ ($a$) is
the inter- (intra-) layer unit cell dimension,
$e$ is the electronic charge,
and the angle brackets denote a normalized
average over the Fermi surface.
For simplicity, we take the interlayer inelastic scattering
to have an Einstein spectrum with frequency $\Omega_0$ but
keep the wave-vector dependence in the impurity-
($V_{\bf q}$) and boson-assisted ($g_{\bf q}$) hopping amplitudes.

The calculation of the penetration depth in the superconducting state
may be performed either by noting that the $c$-axis critical
current $\propto 1 / \lambda_c^2$ in layered
superconductors\cite{Bulaevskii} and computing the critical
current in the standard way,\cite{TMahan,RL} or by calculating
the optical conductivity and extracting the penetration depth
from the imaginary part.\cite{RK}
In either case, one obtains
\begin{equation}
\frac{1}{\lambda_c^2} =
  \frac{32 \pi e^2 d}{\hbar^2 c^2 a^2} \,
  \frac{T^2}{N_{\|}^2} \sum_{k_l k'_{l'}}
  t^2 (k_l, k'_{l'}) \, F (k_l) F (k'_{l'}),
\end{equation}
where $c$ is the speed of light, $T$ is the
temperature ($k_B$ = 1),
$N_{\|}$ is the number of lattice sites in a single
layer, $k_l = ({\bf k}, i\omega_l)$, $F (k_l)$ is
the Gor'kov propagator,\cite{AGD,AM} and $t^2 (k_l, k'_{l'})$ is
the generalized interlayer hopping matrix element.
As with the conductivity, this matrix element is the sum of the
matrix elements due to direct scattering,
$t^{\rm 2,direct} (k_l, k'_{l'}) =
  \frac{N_{\|}}{T} \, \delta_{\bf k,k'} \, \delta_{l,l'} \;
  t_{\bot}^2$;
impurity-assisted scattering,
$t^{\rm 2,imp} (k_l, k'_{l'}) =
  \frac{1}{T} \delta_{l,l'} \, \left| V_{\bf k - k'} \right|^2$;
and boson-assisted scattering,
$t^{\rm 2,inel} (k_l, k'_{l'}) =
  \left| g_{\bf k - k'} \right|^2 D (k_l - k'_{l'})$.
In the last term, $D (k_l)$ is the propagator for the
interlayer boson.
We note that all three mechanisms produce a contribution
to the penetration depth.
This result is not surprising, given the close similarity between
incoherently coupled layers and tunnel junctions:
the direct and impurity-assisted hopping processes formally resemble
the processes considered in conventional SIS
junctions,\cite{Cohen,AB,Duke,TMahan}
and boson-assisted hopping has been known to contribute to
tunneling in superconducting junctions in both the
quasiparticle\cite{Kleinman} and Josephson\cite{Duke} channels
for some time.

To compute the penetration depth from these formulae, we make
several simplifying assumptions.
First, we use the standard BCS form for the Gor'kov propagators
and perform the sums over ${\bf k}$ in the usual way by restricting
the wave vectors of the self-energies and matrix elements to the
Fermi surface and then integrating the remaining energy dependence
from $-\infty$ to $+\infty$.
We reiterate that the weak interlayer coupling allows us to perform
our calculations to second order in the interlayer hopping amplitudes,
and this implies that inter-layer scattering effects which
act to reduce $1 / \lambda_c^2$ are of higher order
and can be neglected.\cite{RL}
Second, we make the reasonable approximation that the
layers are identical.
Third, we account for the anisotropy of the hopping processes
by taking the impurity scattering matrix element
$\left| V_{\bf k-k'} \right|^2$ to have a Lorentzian form with
a maximum at $\bf k=k'$ and a half width $\delta k / k_F$ = 0.01
and by assuming that the wave-vector structure of
$\left| g_{\bf k-k'} \right|^2$ is such that the boson-assisted
processes contribute to both the resistivity and the penetration
depth with the same strength.
The qualitative trends in $\rho_c$ and $\lambda_c$
we discuss do not depend on these choices;
see Ref.~\onlinecite{RK} for details.

{}From these relations,
we see that $\rho_c = 1 / \sigma_c$
is determined by the same parameters as $\lambda_c$.
We are therefore able to connect the two quantities and
examine the qualitative predictions of our theory.
Fig.~\ref{fig:lambdac} shows the resulting curves for
$\rho_c (T)$ and the corresponding $c$-axis penetration
depth ratio for both $s$- and $d$-wave pairing in several
limiting cases of our model.

If only direct hopping is present [cf. Fig.~\ref{fig:lambdac}(a)],
only those terms corresponding to wave function overlap
contribute to the resistivity and the penetration depth.
This case may be relevant for materials like fully oxygenated
YBCO, which is one of the least anisotropic cuprates
($\rho_c (T_c) / \rho_{ab} (T_c) \sim$ 10-10$^2$ compared to
$\rho_c (T_c) / \rho_{ab} (T_c) \sim$ 10$^5$ for BSCCO\cite{Cooper})
and therefore potentially the least incoherent.
In this instance, $\rho_c$ reflects the temperature dependence,
though not the magnitude, of the $ab$-plane resistivity:
$\rho_c \propto \rho_{ab} \propto T$ [cf. Eq.~(\ref{eq:o_direct})].
The resulting penetration depth is also marked by this
near-coherence and has the same temperature dependence for
{\it both} the $ab$- and $c$-axis directions, regardless of
pairing symmetry.

The case of assisted hopping, where only the impurity- and
boson-assisted processes contribute to the $c$-axis transport,
is shown in Fig.~\ref{fig:lambdac}(b).
Materials like BSCCO with $c$-axis mean free paths much less than
the lattice spacing are expected to be close to this limit.
$\rho_c$ in this case is marked by a negative temperature
derivative which gives rise to a semiconducting temperature
dependence above $T_c$.
This upturn in the resistivity is due to the freezing-out of
the inelastic interlayer scattering at low $T$, which inhibits
$c$-axis transport and therefore increases the resistivity.
At lower $T$, however, the impurity scattering acts to
limit the conductivity, and $\rho_c (T)$ saturates.
The $c$-axis penetration depth is also modified by the incoherent
transport and becomes distinct from the $ab$-plane result.
For both pairing symmetries, the $c$-axis penetration depth
ratio is larger than the $ab$ plane penetration depth ratio
at all temperatures.
In particular, the $d$-wave penetration depth ratio
$\lambda_{c}^{2}(0)/\lambda_{c}^{2}(T)$ resembles the $s$-wave
case, although the low-temperature behavior is still a power law.
In addition, the difference between in-plane and $c$-axis penetration
depth ratios is much more pronounced for $d$-wave pairing than for
$s$-wave paring (see the upper inset in Fig.~\ref{fig:lambdac}(b)).
Moreover, the temperature dependence in the $d$-wave case is close
to, and may be experimentally indistinguishable from, the
($s$-wave) Ambegaokar-Baratoff\cite{AB} form.

In the intermediate case, all processes contribute to $c$-axis
transport [Fig.~\ref{fig:lambdac}(c)].
Compounds such as de-oxygenated YBCO may belong to this class.
As in the assisted hopping case, $\rho_c$ looks semiconducting
at high temperatures.
At low temperatures, however, the in-plane scattering rate is
reduced, and this leads to a low-$T$ conductivity dominated by
direct hopping processes [cf. Eq.~(\ref{eq:o_direct})].
The net result is a peak in $\rho_c$ which may lie below $T_c$.
In contrast to the resistivity, the penetration depth ratio shows
no new behavior in this case, but is mid-way between the direct
result, where both $ab$- and $c$-axis penetration depth ratios
are the same, and the assisted hopping result, where they are
considerably different.
Note especially that the low-temperature $c$-axis penetration
depth ratio for $d$-wave pairing is clearly linear in $T$ but
with a much smaller slope than its in-plane counterpart.

Fundamentally, the behavior discussed above results from the
different characteristic temperature dependences
that arise when the interlayer coupling is very weak (incoherent)
as opposed to when it is strong (coherent).
In a material with very weak interlayer coupling,
quasiparticle transport between adjacent layers is analogous
to tunneling in NIN or SIS junctions.
This tunneling is mediated by both elastic are inelastic
scattering and is therefore associated with a semiconducting
$c$-axis resistivity in the normal state and a very low
slope of the $c$-axis penetration depth ratio at low temperatures.
The small low-temperature slope of $\lambda_c^{-2} (T)$
is a special feature of specular Josephson tunneling and occurs
in either $s-$\cite{AB} or $d$-wave\cite{RK} superconductors.
On the other hand, in materials with stronger interlayer coupling,
quasiparticle transport is nearly coherent and so
quasi-three-dimensional results obtain:
$\rho_c$ is metallic, and $\lambda_c^2 (0) / \lambda_c^2 (T)$
possesses a larger low-$T$ slope for $d$-wave pairing
relative to the weakly coupled case.

Empirically, therefore, one should associate a reduction in
the low-$T$ slope of the $c$-axis penetration depth ratio in
{\it either} $s$- or $d$-wave superconductors with a decrease
in the interlayer coupling and a semiconducting temperature
dependence of $\rho_c$.
Our theory thus accommodates the currently available data in
{\it both} the intra- and interlayer directions despite their
very different low-temperature slopes.
Moreover, this consistency suggests that the cuprates may
indeed be $d$-wave superconductors.
While our theory can explain the qualitative features of
the present experimental data, these data are incomplete;
further systematic experiments on different cuprates and for
particular cuprates with different stoichiometries are clearly
called for to further test the trends reported in this paper.
In addition, further theoretical effort is required to understand
the relationship of our theory to others in the literature
and to provide a way of distinguishing them experimentally.

Among these other theories of $c$-axis coupling, the ``confinement''
approach\cite{AZ} is worth discussing further, since we come
to similar conclusions despite vastly different starting assumptions.
In contrast to our Fermi-liquid-based approach,
the confinement theory asserts that each CuO$_2$ layer is a
spin-charge separated tomographic Luttinger liquid.\cite{AZ}
Nevertheless, the expressions for the interlayer transport
which arise in this model are similar to those in this work
with two important differences:
(1)~the Green's functions used in the calculation
correspond to a Luttinger and not a Fermi liquid, and
(2)~inelastic scattering is an intrinsic part of the interlayer
transport process and is not extrinsic to the layers as in,
for example, the boson-assisted hopping term in our theory.
Despite these differences, both theories obtain a semiconducting
$T$ dependence of $\rho_c$\cite{AZ} and a small
low-$T$ slope of the penetration depth ratio,\cite{Wheatley}
in agreement with experiments.
This comparison suggests that models of $c$-axis coupling in the
cuprates should view interlayer transport as a tunneling process
with a strong inelastic component, but the detailed
origin of this effect may not be readily extracted from
the currently available experiments.

The authors would like to thank D. A. Bonn and S. Kamal
for making their penetration depth data available to us,
and T. Timusk and S. Anlage for valuable discussions.
This work was supported by the National Science Foundation
(DMR-9123577 and DMR 91-20000, through the Science and
Technology Center for Superconductivity) and by
NASA Grant No. NAG3-1395.

\end{document}